# Ultrathin wave plates based on bi-resonant silicon Huygens' metasurfaces

Zhiyuan Fan[1] and Gennady Shvets[1,*]

[1]School of Applied and Engineering Physics, Cornell University, Ithaca, NY 14853
*Corresponding author: gshvets@cornell.edu

**Abstract**

**An all-dielectric Huygens metasurface supporting electric and magnetic resonances is a promising platform for a variety of optical applications that require a combination of extremely high transmission and broad-range control of the phase of the transmitted light. A highly efficient wave plate is one example of such an application. In combination with high spectral selectivity and strong optical energy concentration, such phase plates are desirable for precision sensing and high efficiency nonlinear optics. We propose a novel approach to realizing such ultra-thin optical plates: an anisotropic Fano-resonant optical metasurface (AFROM) employing the combination of a spectrally sharp electric and a relatively broadband magnetic resonance. The phase shift coverage approaching $2\pi$ is achieved through judicious choice of the geometric parameters of a complex unit cell. A new methodology based on eigenvalue simulations of leaky magnetic/electric resonances enables rapid computational design of such metasurfaces.**

## 1. INTRODUCTION

Polarization is a fundamental degree of freedom of a light wave, along with its intensity and phase. The ability to manipulate the state of polarization (SOP) is crucial for a variety of optical application such as glare reduction, surface characterization, detection of chiral molecular enantiomers, and even improved laser machining [1], to name just a few. Changing light's polarization requires changing the phase relationship between the two orthogonal components of light's electric field components, i.e. $E_x$ and $E_y$ for the light waves propagating in the $z-$direction. This is usually accomplished by wave plates (also known as phase retarders) that employ natural birefringent materials with different refractive indices $n_x$ and $n_y$ for the two linear polarizations.

The need for novel approaches to polarization manipulation of light is particularly urgent in the mid-wave infrared (mid-IR) spectral range, where fewer standard phase-retarding optical components are available than in the visible rage despite the rapidly growing interest in mid-IR applications such as free-space optical communications [2, 3], remote gas sensing [4], and Stokes polarimetry [5, 6, 7].

Some of the most common polarization conversion applications involve transforming linearly polarized (LP) light into either left- or right-hand circularly polarized (LCP or RCP) states (quarter-wave plates or three-quarter-wave plates), or rotating the LP light by 90° (half-wave plates). For example, LP-to-CP conversion is important in high-power applications that involve laser processing of materials (e.g., hole drilling), as well as for suppressing specular reflections. Similarly, half-wave plates, in combination with polarizing beam splitters, can be used for variable light attenuation.

One obvious limitation of standard wave plates is their thickness. The typically small refractive index contrast $\Delta n = |n_x - n_y|$ cases the wave plates thickness $h$ to be larger than the targeted wavelength of light $\lambda$. Metamaterials and metasurfaces [8, 9, 10] are natural candidates for enhancing the effective birefringence and, therefore, shrinking the size of the wave plates. A number of distinct types of metasurfaces (reflective, transmissive, metallic, or dielectric) have been recently deployed for polarization control and conversion [8, 9, 10, 11, 12, 13]. While plasmonic metasurfaces can be made extremely thin (tens of nanometers), they have limited conversion efficiency [10] and low laser damage threshold. Many of the reported birefringent dielectric metasurfaces [14, 15, 16, 17, 12] have the thickness $h \sim \lambda$. This presents a particular challenge to mid-IR applications that would require fairly thick metasurfaces.

In addition to their sub-wavelength dimensions, several key requirements must be met by polarization-controlling transmitting metasurfaces in order to make them appealing to the broadest range of applications. First, many high-power applications require extremely high transmittance $T$ because even the smallest amount of loss or reflection can, correspondingly, damage the metasurface or destabilize the high-power laser system. For any metasurface deployed for such applications, the starting design point must be $T = 1$ at the operating wavelength because the inevitable imperfections of the photonic structure will reduce the transmission. Stringent transmittance requirements rule out promising plasmonic metasurface concepts [11, 10, 18, 19] because of their Ohmic loss at optical frequencies.

Second, polarization-converting metasurfaces have to be extremely anisotropic in order to provide the largest possible range of phase differences $\Delta\phi = \phi_y - \phi_x$ between the two polarizations. For example, it requires the phase difference coverage of at least $\pi/2 < \Delta\phi < 3\pi/2$ range to convert LP light into either one of the two CP polarizations, or to rotate its polarization by 90°. Ideally, an even broader range of $\Delta\phi$ (close to $2\pi$) would be desirable to enable the transformation of LP light into arbitrarily elliptically polarized one. While many successful metasurface designs were shown to produce large phase retardations for one or both polarization [13, 20, 21, 22, 23], the challenge remains to produce a large range of $\Delta\phi$, all the while maintaining subwavelength thickness of the metasurface combined with ultra-high transmittances $T_{x,y}$ for both polarizations.

One approach to achieving such high transmittance is to utilize a recently introduced [23, 24] concept of a dielectric Huygens (i.e. perfectly transmitting) metasurface that relies on combining a pair of crossed electric dipole active (E-) and magnetic dipole active (M-) modes that are simultaneously excited by an incident light wave. Such structures can be based on silicon resonant meta-atoms that combine the benefits of CMOS compatibility with the ease of single-layer lithographic fabrication. Si-based metasurfaces have negligible Ohmic losses at mid-IR, while the reflection is suppressed due to the simultaneous excitation of the electric and magnetic dipole active modes [25, 26, 27, 28]. It has been demonstrated [20, 23] that matching of the frequencies and lifetimes of the E- and M-modes of Si disk arrays is sufficient for a unity transmission with a nearly-$2\pi$ phase change. However, the proposed designs are isotropic and do not lend themselves to making Huygens phase plates.

## 2. SUMMARY OF THE RESULTS

In this article, we show a new approach to designing Huygens wave plates using anisotropic metasurfaces comprised of silicon meta-molecules with broken mirror symmetry. This asymmetry enables resonant excitation of high quality factor (high-Q) E- and M-modes by only one linear polarization of light (e.g., the y-polarization). The orthogonal (non-resonant) polarization of light does not significantly interact with the metasurface, so its transmission phase $\phi_x$ is nearly frequency-independent while its transmittance $T_x \approx 1$.

Using a coupled mode theory (CMT) [29, 30, 31, 32, 33], we show that for any set of complex-valued resonant frequencies $\widetilde{\omega}_E$ and $\widetilde{\omega}_M$ of the E- and M-modes, respectively, there exists a real-valued Huygens frequency $\omega \equiv \omega_H$ for which the transmittance of the resonant polarization is unity: $T_y(\omega_H) = 1$. The phase $\phi_y$ of the transmitted resonant polarization at the Huygens frequency can assume any value in the $0 < \phi_y < 2\pi$ range depending on the relative spectral positions of the E- and M-resonances. Therefore, by varying the resonant frequencies of the electric and magnetic dipole resonances of a meta-molecule with respect to each other, we can achieve near-100% transmittance with an arbitrary phase difference between the two polarizations, thus enabling a wide variety of functional phase plates. The reductive description of the Huygens metasurface in terms of its two resonances and their respective frequencies $\omega_{E(M)}$ and radiative lifetimes $\gamma_{E(M)}^{-1}$ enables rapid search for the geometry of the metasurfaces that provide the desired SOP of the transmitted light.

While the described wave plate is inherently narrowband due to its resonant nature, it has an important advantage of deeply subwavelength ($h < \lambda/4$) thickness which is important for a variety of mid-IR applications that do not require broadband operation (e.g., remote gas sensing). Moreover, as we show in the Supplemental Materials (SM) section, the bandwidth can be significantly broadened by straightforward design modifications. Transmissive metamaterial pixels with different Huygens frequencies can also be spatially multiplexed because their sharp resonant response does not rely on long-range interaction between metamolecules [34].

## 3. COUPLED MODE THEORY (CMT) OF A RESONANT WAVE PLATE

In the framework of the CMT, the general description of the excitation of the two dipole-active modes by the $y-$ polarized light (which is assumed to be the resonant linear polarization) is given by the following equation:

$$\begin{pmatrix} a_E \\ a_M \end{pmatrix} = -i \begin{pmatrix} \omega - \widetilde{\omega}_E & 0 \\ 0 & \omega - \widetilde{\omega}_M \end{pmatrix}^{-1} \begin{pmatrix} \alpha_E E_y \\ \alpha_M H_x \end{pmatrix}, \quad (1)$$

where the two modes are, respectively, an electric mode resonating at $\omega_E$ and a magnetic mode resonating at $\omega_M$. Here $\widetilde{\omega}_{E(M)} = \omega_{E(M)} - i\gamma_{E(M)}$ are the complex-valued frequencies of the E- and M-modes, respectively. Neglecting the non-radiative (Ohmic) losses, the inverse lifetimes of the modes are directly related to the modes' far field coupling coefficients $\alpha_{E(M)}$ according to $\gamma_{E(M)} \equiv \alpha_{E(M)}^2$. Without loss of generality, we assume that the metasurface is surrounded by air on both sides and that the incident wave is planar ($E_y = H_x$).

To simplify the analysis, we further assume that that the electric dipole mode is much more narrow-band than the magnetic dipole mode, i.e. $\gamma_E \ll \gamma_M$. The reported results do not rely on the above relationship between the lifetimes of the modes, which is used here to simplify the expression for the Huygens frequency. One specific design that is the focus of this paper is shown in Fig.2. Due to the mirror symmetry of the metamolecules, $x-$ and $y-$axes form the principal axes. Thus, no cross-polarization conversion between $x-$ and $y-$polarized light takes place, and the E- and M-modes shown in Figs. 2(b,c) are not excited by the $x-$polarized light. The spectrally sharp E-mode emerges as a result of Fano interference [34, 35, 36] of the "dark" electric quadrupole mode with a "bright" electric dipole mode [34].

Under these assumptions, an explicit general form of the complex-valued transmission coefficient $t_y \equiv t(\omega)$ (which includes the phase and transmittance information: $t = \sqrt{T}e^{i\phi}$, where $\phi(\omega) \equiv \phi_y$ and $T(\omega) \equiv T_y$) can be obtained [23, 37, 38]:

$$t(\omega) = 1 - \frac{i\gamma_M}{\omega - \widetilde{\omega}_M} - \frac{i\gamma_E}{\omega - \widetilde{\omega}_E}. \quad (2)$$

Huygens transmission can be understood from Eq. (2) by observing that there always exists a frequency $\omega_H$ such that $t(\omega_H) = e^{i\phi_H}$. This frequency, further referred to as the Huygens Frequency (HF), is given by the following simple expression:

$$\omega_H(\widetilde{\omega}_M, \widetilde{\omega}_E) = \frac{\gamma_E \omega_M - \gamma_M \omega_E}{\gamma_E - \gamma_M} \approx \omega_E, \quad (3)$$

where the final simplification is obtained in the $\gamma_E \ll \gamma_M$ limit.

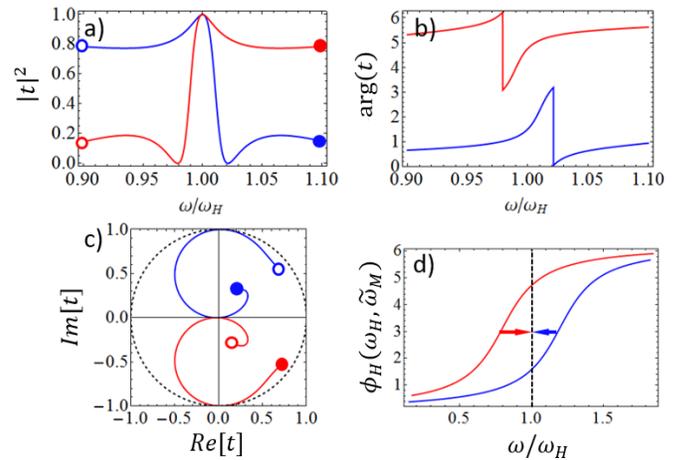

**Fig. 1.** Light transmission through a phenomenological bi-resonant metasurface: (a) amplitude, (b) phase, (c) complex-valued transmission coefficient. The metasurface is defined by Eqs.(1,2); two examples are considered: $\widetilde{\omega}_M/\omega_H = 1.2 - 0.2i$ (blue curves) and $\widetilde{\omega}_M/\omega_H = 0.8 - 0.2i$ (red curves). The electric resonance is assumed to be narrow-band ($\gamma_E = 0.01\omega_E$); frequencies are normalized to the Huygens frequency $\omega_H$ kept the same for both examples. Hollow (solid) circles correspond to $\omega = 0.9\omega_H$ ($\omega = 1.1\omega_H$). (d) Transmission phases $\phi(\omega_H, \widetilde{\omega}_M^{(1,2)})$ at perfect transmission from Eq.(4). Dashed line: the Huygens frequency. Black dashed line indicates the necessary condition for HSA $\omega_H \approx \omega_E$. The arrows indicate the fractional frequency detuning $\Delta = (\omega_M - \omega_E)/\omega_E$ between electric and magnetic modes.

The transmittance spectra $T(\omega)$ are plotted in Fig. 1(a) for two examples (blue and red solid lines) corresponding to $\widetilde{\omega}_M^{(1)}/\omega_E^{(1)} = 1.2 - 0.2i$ (blue curves) and $\widetilde{\omega}_M^{(2)}/\omega_E^{(2)} = 0.8 - 0.2i$ (red curves). The corresponding frequencies of the electric resonances, $\omega_E^{(1)}$ and $\omega_E^{(2)}$, were selected in such a way that unitary transmission is achieved at the same HF: $\omega = \omega_H$. It follows from Eq. (3) that, in the limit of a high-Q electric resonance, $\omega_H \approx \widetilde{\omega}_E^{(1)} \approx \widetilde{\omega}_E^{(2)}$ is very close to that of the electric resonance, but far from the magnetic resonance frequencies satisfying $\left|\omega_M^{(1)} - \omega_M^{(2)}\right| \sim \gamma_M^{(1,2)} \gg \gamma_E$.

Under the perfect transmission assumption, the Huygens phase $\phi_H \equiv \phi(\omega = \omega_H, \widetilde{\omega}_M)$ is given by
$$\phi_H(\omega, \widetilde{\omega}_M) = -2 \tan^{-1}\left(\frac{\gamma_M}{\omega - \omega_M}\right), \quad (4)$$
where the interpretation of Eq.(4) is as follows: for any values of $\omega$ and $\widetilde{\omega}_M$, there exists a complex-valued electric resonance frequency $\widetilde{\omega}_E$, calculated from Eq. (3), such that $\omega = \omega_H$. For those parameters, the transmission is perfect, and the transmitted phase is given by Eq. (4) at $\omega = \omega_H$. Note that the zeros of the transmission amplitudes $T^{(1,2)}(\omega)$ shown in Fig. 1(a) occur at $\omega_E^{(1)}$ and $\omega_E^{(2)}$, respectively.

The plots of the transmitted phases $\phi^{(1,2)}(\omega)$ in Fig.1(b) demonstrate that the transmission phases $\phi^{(1)}(\omega) \neq \phi^{(2)}(\omega)$ for all frequencies (including $\omega = \omega_H$) in the two examples corresponding to the M-modes' resonance on the blue (blue curve) and the red (red curve) sides of the E-mode's resonance. Specifically, it can be observed from Fig.1(c) that $\phi_H \equiv \phi_H^{(1)} = \pi/2$ for the first example, while $\phi_H \equiv \phi_H^{(2)} = 3\pi/2$ for the second example. Therefore, the phase shift of 180° can be accomplished by sweeping the frequency of the magnetic resonance across a sharp electric resonance that determines the Huygens frequency.

Because perfect transmission is achieved only for $\omega = \omega_H$, the intersection of the $\phi_H(\omega, \widetilde{\omega}_M^{(1)})$ and $\phi_H(\omega, \widetilde{\omega}_M^{(2)})$ plots (solid line) and the $\omega = \omega_H$ (dashed line) in Fig.1(d) recovers the Huygens phases, $\varphi_H^{(1)}$ and $\varphi_H^{(2)}$, of the perfectly-transmitted light. Clearly, by choosing a family of structures with $\omega_M^{(1)} < \omega_M^{(j)} < \omega_M^{(2)}$, we can provide the phase shift coverage of $\pi/2 < \varphi_H^{(j)} < 3\pi/2$. Note that the difference between $\omega_E^{(1)}$ and $\omega_E^{(2)}$ is of order $\gamma_E^{(1,2)} \ll \gamma_M^{(1,2)}$, so the frequencies of the two E-modes are indistinguishable in Fig.1(d).

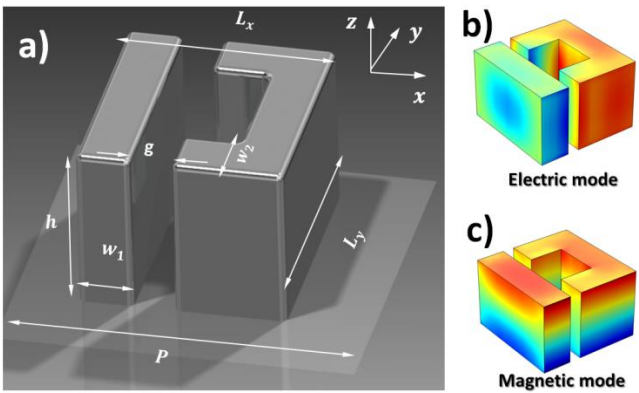

**Fig. 2.** A specific implementation of dielectric anisotropic Fano-resonant optical metasurface supporting magnetic and electric resonances. (a) A schematic and geometric parameters' definitions of a unit cell of a Si Huygens metasurface. Color-coded $E_y$ for an electric (b) and magnetic (c) resonances. Red: $E_y > 0$, blue: $E_y < 0$.

The phenomenological examples analyzed in Fig.1 show that, by sweeping the frequency of the magnetic resonance across that of the electric one by approximately one radiative linewidth of the former, we can achieve more than 180 degrees of phase shift. While the idea of bringing the magnetic and electric resonances close to each other has been discussed in the past [20, 23], the concept of swapping the spectral order the E-mode and M-mode for obtaining arbitrary phase shifts is discussed here for the first time.

The example presented above points to a viable and so far unexplored strategy for designing a lossless Huygens metasurface capable of imparting an essentially arbitrary phase shift to transmitted light. Our strategy involves starting with a sufficiently complex (see, for example, Fig.2) unit cell that supports a sharp electric and a broad magnetic resonances, and then perturbing the dimensions of the unit cell so as to keep the frequency of the electric resonances fixed at $\omega_E$ while significantly varying the frequency of the magnetic resonance $\omega_M$. If the topology of the unit cell possesses a sufficient number of degrees of freedom, then this strategy can be easily implemented using the following straightforward procedure (see SI for more details). For each $i$'th set of unit cell's geometric parameters, we numerically calculate the corresponding frequency triplets $\mathbf{\Omega}^{(i)} \equiv (\omega_E^{(i)}, \omega_M^{(i)}, \gamma_M^{(i)})$, and then down-select only those members of the set that correspond to the same electric dipole frequency $\omega_E^{(i)} \equiv \omega_H$.

## 4. EXAMPLE OF A SPECIFIC DESIGN: FANO-RESONANT ANISOTROPIC METASURFACE

One specific design of a dielectric-based AFROM satisfying the $\gamma_E \ll \gamma_M$ condition is shown in Fig.2, where the unit cell is comprised of two unequal (straight and C-shaped) Si antennas. The asymmetry between the antennas gives rise to the high-Q E-mode shown in Fig.2(b). The finite thickness $h$ of the metasurface is responsible for the low-Q M-mode shown in Fig.2(c). The high-Q electric dipole mode emerges as a result of a mirror-symmetry breaking that hybridizes the "dark" electric quadrupole mode with a "bright" electric dipole mode [34]. The new hybridized electric mode is modeled here using a very small radiative decay constant $\gamma_E$. This modes' hybridization can be interpreted as their Fano interference [34, 35].

For the unit cell shown in Fig.2, the geometric parameters set is given by $\mathbf{i} \equiv \{L_x, L_y, w_1, w_2, g\}$, where the dimensions are labelled in Fig.2(a). The corresponding frequency vectors $\mathbf{\Omega}^{(i)}$ are obtained from the eigenvalue simulations using the COMSOL Multiphysics® [39] software. The simulations also reveal the field distributions of $E_y$ corresponding to the electric (shown in Fig. 2(b)) and magnetic (shown in Fig.2(c)) resonances. The magnetic mode has a magnetic dipole moment in the $x$-direction, and the electric mode has an electric dipole moment in the $y$-direction. The AFROM's optical response to $x$-polarized light can be neglected because of the high aspect ratio of each other two antennas.

We numerically confirmed that the assumption of the sharpness of the electric resonance ($\gamma_E \ll \gamma_M$) is indeed satisfied for a very broad range of geometric parameters. In the rest of the paper we keep the metasurface thickness $h$ fixed because of the practical convenience of fabricating planar structures vs the three-dimensional ones. Therefore, in what follows all frequencies are normalized to $\omega_{0h} = 2\pi c/h$. In addition, to ensure that a metasurface does not become too large or too small, the unit cells are arranged in a square lattice with a fixed period $P$. Through the rest of the paper, we assume $P = 3\mu m$ and $h = 1.2\mu m$, and further confine the parameters space to the mid-infrared (mid-IR) spectral range, so that the normalized frequencies $\omega/\omega_{0h} \sim 0.25$, which implies that the thickness of the AFROM is of order $\lambda_H/4$ of an operating wavelength $\lambda_H$ (i.e. it is strongly sub-wavelength).

To illustrate that the five-component geometric parameters vector $\mathbf{i}$ provides sufficient freedom for keeping the electric resonance frequency constant while sweeping the magnetic resonance, we have reduced it even further by fixing the values of $L_y$, $w_1$, and $w_2$ (listed in the caption to Fig.3), and varied the remaining two parameters: $L_x$ and $g$. The results of finding the electric/magnetic modes' eigenvalues as a function of these two design parameters are presented in Fig.3, where the constant frequency contours $\omega_E(L_x, g)$ (solid lines) and $\omega_M(L_x, g)$ (dashed lines) are plotted.

In addition, the calculated HP from Eq.(4) is presented in the same figure as a color-coded $\phi_H(L_x, g)$ map. For example, the $\omega_E = 0.265\omega_{oh}$ contour is intercepted by all four $\omega_M = const$ contours. Based on the color-coded HP, we conclude that a phase difference $\delta\phi_H \approx 2\, rad$ can be obtained by varying these two structure parameters while keeping the electric resonance frequency (and, to the lowest approximation according to Eq.(3), the Huygens frequency $\omega_H \approx \omega_E$) constant. Of course, all five dimensions can be independently varied. Therefore, much larger tunability of the Huygens phase can be achieved at a given operating frequency.

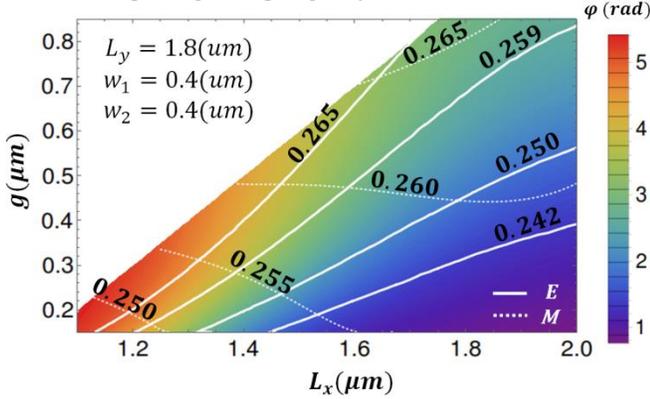

**Fig. 3.** Resonance frequencies (contours) and color-coded Huygens phase shift of air-suspended all-Si AFROM. White solid lines: $\omega_E(g, L_x) = $ const. contours, white dotted lines: $\omega_E(g, L_x) = $ const. contours. The complex-valued resonant frequencies were obtained from eigenvalue simulations using COMSOL Multiphysics software. The frequencies are normalized to $\omega_{0h}$ and marked on the constant frequency contours. Fixed dimensions: $P = 3\mu m, h = 1.2\mu m, w_1 = w_2 = 0.4\mu m, L_y = 1.8\mu m$. Huygens phase is obtained from Eq.(4). Dielectric permittivity of Si: $\epsilon_{Si} = 12$. Geometric parameters are defined in Fig.2.

Additional qualitative conclusions can be drawn from Fig.3. First, if relative variation of $g$ and $L_x$ are to be kept reasonably small (as exemplified by the ranges shown in Fig.3), then there is an approximate relationship between the thickness of the metasurface and the operating wavelength that provides for the largest variance of $\delta\phi_H$ with respect to the variations of $g$ and $L_x$. Specifically, by inspecting Fig. 3, we can deduce that the optimal operating frequency range is around $\omega/\omega_{0h} \sim 0.25$, which corresponds to metasurface thickness of order $\lambda/4$. Thinner metasurfaces would require much more extreme variations of all geometric parameters, and will not be considered further in our designs. We note that most of the earlier proposed phase plates designs that did not utilize electric and magnetic resonances either had to relax the transmission requirements [13], or to employ thicker metasurfaces [12].

More detailed simulations that include the variation of all five geometry parameters are presented in the SI section, and illustrated in Fig.S1. The $0.24 < \omega/\omega_{oh} < 0.28$ frequency range is found to be optimal for constructing a family of metasurfaces that spans the $\delta\phi_H \approx 5\, rad$ range of Huygens phases. The calculation is based on the following two-step procedure: (i) extracting the complex-valued eigen frequencies $\widetilde\omega_E$ and $\widetilde\omega_M$ of the electric and magnetic resonances for the metasurface shown in Fig.2, and (ii) substituting them into Eq.(4) to determine the Huygens phase $\phi_H(\omega = \omega_H, \widetilde\omega_E, \widetilde\omega_M)$. The effectiveness of this approach was verified using driven (i.e. with a normally incident EM wave) COMSOL simulations. Such driven simulations extract the phase of transmitted wave as function of the frequency of incident light. Phase coverage of $\delta\phi_H \approx 2\pi$ phase is obtained for transmission efficiencies $T > 95\%$ as shown in Fig.S3 of the SI.

So far our discussion was limited to calculating the phase shift $\phi_y \equiv \phi_H$ of the $y$-polarized light. However, the design of the AFROM shown in Fig. 2 is deliberately anisotropic in order to produce a polarization-transforming phase plate. We now proceed to demonstrate their performance by including their non-resonant response to $x$-polarized light. Therefore, the search in the space of the five-component geometric parameters recipe needs to take into account the phase $\phi_x$ of the $x$-polarized light. We carried out the eigenmode-based search for those structures that support high-Q $y$-polarized E-modes in the desired frequency range, and conduct full wave simulation that accepts only those geometries that satisfy the following conditions: (a) $T_x, T_y > 0.95$, and (b) $\Delta\phi = \phi_y - \phi_x$ corresponds to the desired polarization of the transmitted light.

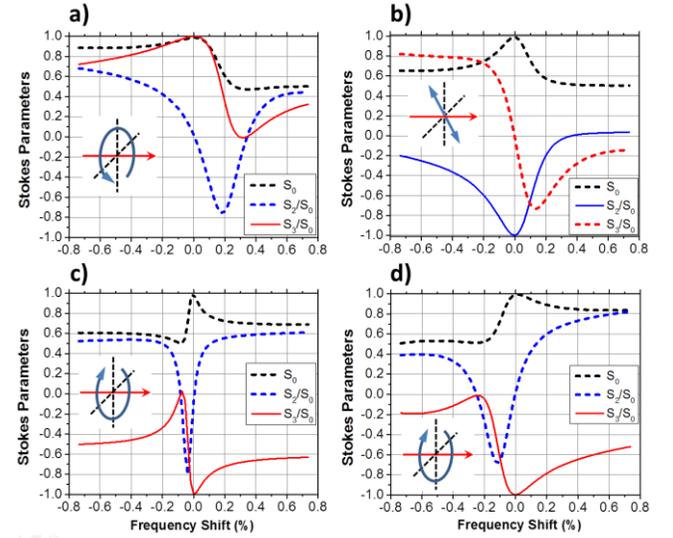

**Fig. 4.** Performance of four Si AFROM designs characterized by the Stokes parameters of the transmitted light: air-bridged metasurfaces operating as quarter-wave plates (a), half-wave plates (b), and three quarter-wave plates (c). (d) Same as (c), but with a quartz substrate and air superstrate. Horizontal axes: frequency shifts from $\omega_H = 0.272\omega_{oh}$. Insets: the polarization state of transmitted beams. The incident beam is linearly polarized at $45°$ with respect to the principal axes. Thickness of the metasurface: $h = 1.2\mu m$. Huygens wavelength: $\lambda_H = 2\pi c/\omega_H = 4.41\mu m$. Other geometric sizes: see Table S3 of the SI.

The performance characteristics of three such designs that transform linearly polarized incident (LP) light ($\mathbf{E}_{inc} = \mathbf{e}_x + \mathbf{e}_y$) into left handed circularly polarized light (LCP: quarter-wave plate), orthogonally polarized LP light (half-wave plate), and right handed circularly polarized light (RCP: three quarter-wave plate), are shown in Figs. 4(a-c), respectively. The geometric dimensions of the three wave plates, which operate at $\omega_H = 0.272\omega_{oh}$, are listed in Table S3 of the SI. Specifically, the three relevant Stokes parameters $2S_0(\omega) \equiv (T_{+45°} + T_{-45°}) \equiv (T_{RCP} + T_{LCP})$ (black dashed line), $S_2/S_0 \equiv (T_{+45°} - T_{-45°})/(T_{+45°} + T_{-45°})$ (blue dashed line), and $S_3/S_0 \equiv (T_{RCP} - T_{LCP})/(T_{RCP} + T_{LCP})$ (red solid line) are plotted in Figs. 4(a-c). By definition of the Huygens frequency, $S_0(\omega_H) = 1$, as shown in Figs.4(a-c) for all three wave plate designs. The effect of a low-index substrate (neglected in Figs. 3(a-c), where air-bridged metasurfaces were assumed) is included in a representative design of a three quarter-wave design shown in Fig.4(d). The Si AFROM is assumed to be placed on a quartz substrate ($\epsilon_{SiO_2} = 2.25$). Note that the total transmission is still very close to perfect, and the normalized Stokes parameter $S_3 \approx -1$ signifying that the transmitted light is LCP.

## 5. CONCLUSIONS

We have demonstrated a new approach to designing perfectly transmitting (Huygens) phase plates based on utilizing two electromagnetic resonances of a high-index anisotropic metasurface. The examples presented in this article demonstrated that essentially any polarization state of light can be obtained from the incident LP wave. Although many high-intensity laser applications do not require broadband operation, we note that the specific design that was used as an example in this article has a very narrow spectral width ($Q \sim 500$), However, the same approach of using unequal asymmetric dielectric antennas can be used for designing AFROMs with much lower spectral selectivity ($Q \sim 60$) as shown in Fig. S6 of the SI.

**Funding sources and acknowledgments.** This work was supported by the Office of Naval Research Office Award N00014-17-1-2161.